\def\etal{{\it et al.\thinspace}}

\documentstyle[emulateapj,psfig,apjfonts]{article}

\received{}
\accepted{}
\journalid{}{}
\articleid{}{} 

\lefthead{}
\righthead{}

\begin{document}

\title{Predicting Planets in Known Extra-Solar Planetary Systems II: Testing
for Saturn-mass Planets}

\author{Sean N. Raymond\altaffilmark{1}, Rory Barnes\altaffilmark{1}}

\altaffiltext{1}{Department of Astronomy, University of Washington, Seattle, 
WA, 98195
 (raymond@astro.washington.edu, rory@astro.washington.edu)}

\begin{abstract}
Recent results have shown that many of the known extrasolar planetary systems
contain regions which are stable for massless test particles.  We examine the
possibility that Saturn-mass planets exist in these systems, just below the
detection threshold, and attempt to predict likely orbital parameters for such
unseen planets.  To do this, we insert a Saturn-mass planet into the stable
regions of these systems and integrate its orbit for 100 million years.  We
conduct 200-600 of these experiments to test parameter space in HD37124,
HD38529, 55Cnc, and HD74156. In HD37124 the global maximum of the survival
rate of Saturns in parameter space is at semimajor axis $a$ = 1.03 AU,
eccentricity $e \sim$ 0.1.  In HD38529, only 5\% of Saturns are unstable, and the
region in which a Saturn could survive is very broad, centered on $0.5<a<0.6$,
$e<0.2$. In 55Cnc we find three maxima at ($a$, $e$) = (1.0 AU, 0.02), (2.0
AU, 0.08), and (3.0 AU, 0.17).
In HD74156 we find a broad maximum with $a$ = 0.9-1.2 AU, $e \leq$
0.15.  Several of
these maxima are located in the habitable zones of their parent stars and
are therefore of astrobiological interest.  We suggest the possibility that 
companions may lie in these locations of parameter space, and encourage
further observational investigation of these systems.

 \end{abstract}



\section{Introduction}

There are currently 110 known extrasolar planets, including ten systems
containing two or more planets.  These planets are known to be Jovian both
from their large masses, which range from 0.11 Jupiter masses (HD49674; Butler
\etal 2003) to 17.5 Jupiter masses (HD202206; Udry \etal 2002), and from
their sizes, measured in HD209458 to be 1.27 Jupiter radii (Charbonneau et
al. 2000).  The vast majority of these planets were
discovered by the radial velocity technique, which is sensitive to roughly
3-10 $m \, s^{-1}$(Butler \etal 1996; Baranne \etal 1996).  

All planetary systems must be dynamically stable for at least the
age of their host star.  Recent work by Barnes \& Quinn (2004) suggests that a
large fraction of systems are on the edge of stability: a small change in
semimajor axis $a$ or eccentricity $e$ causes the system to become unstable.
The ``packed planetary systems'' (PPS) hypothesis presented in Barnes \&
Raymond (2004; hereafter Paper 1) 
predicts that all planetary systems are ``on the edge.''  This leads to
speculation that those systems which appear stable may harbor unseen planets
which push them to the edge of stability.  The PPS hypothesis suggests
that if a region exists in a planetary system in which the orbit of a massive
planet is stable, then its presence is likely.

The first paper of this series (Paper 1)
used integrations of massless test particles to map the stability of regions
in certain extrasolar planetary systems in ($a$, $e$) space.  Of the five
systems examined, three (HD37124, HD38529, and 55Cnc) were found to contain
zones between the giant planets in which test particles were dynamically
stable for 5-10 Myr.  Stable regions have been found in $a$ space (assuming
circular orbits) for $\upsilon$ And (Rivera \& Lissauer 2000), GJ876 (Rivera
\& Lissauer 2001) and 55Cnc (Rivera \& Haghighipour 2003). 

In this work we test for the presence of unseen Saturn-mass planets
in four known extrasolar planetary systems: HD37124 (Butler \etal 2003),
HD38539 (Fischer \etal 2003), 55Cnc (Marcy \etal 2002), and
HD74156 (Naef et al 2004).  We choose Saturn-mass planets 
because they lie roughly at the detection threshold for the current radial
velocity surveys (Butler \etal 1996).  The reflex velocity caused by a
Saturn-mass planet at 1 AU on a solar-mass star is 8.5 $m \, s^{-1}$, and scales
with the planet's semimajor axis as $a^{-1/2}$.  For comparison, the smallest
amplitude reflex velocity of any detected planet is 11 $m \, s^{-1}$ (HD1641;
Marcy \etal 2000). Although seven sub-Saturn mass planets have been
discovered as of November 2003 (e.g. Fischer \etal 2003), none has $a >$
0.35 AU.\footnote{Data from http://www.exoplanets.org}

Paper 1 found that no test particles survived in HD74156 for longer than a few
Myr.  However, Dvorak \etal (2003) found orbits stable for test particles
between 0.9 and 1.4 AU. We therefore include HD74156 in our sample.

Table 1 shows the orbital parameters for the four extrasolar planetary systems
we investigate.  Note that the best fit orbital elements for some systems,
especially HD74156c, have changed many times.  We therefore adopt elements as
of a given date, with the knowledge that they may fluctuate. 
In $\S$2 we describe our initial conditions and numerical
method.  We present the results for each planetary system in $\S$3, and
compare these with other work in $\S$4.  We present our conclusions in $\S$5.

\section{Numerical Method}

For each planetary system in Table 1, 200 to 600 values of $a$ and $e$ are
selected at random from within the regions which are stable for test
particles, shown in Table 2.  In the case of HD74156, which has no stable
region, we drew values from the following region: $\Delta a$ = 0.5-1.5 AU,
$\Delta e$ = 0.0-0.2.  For each of these ($a$, $e$) points we assign the new
planet one Saturn mass, an inclination of 0.1$^{\circ}$, and a randomly chosen
mean anomaly.  The longitude of periastron is aligned with the most massive
giant planet in the system.  This assumption helps find more
stable systems, as most of the known planetary systems with ratios of orbital
periods less than 5:1 are found to be librating about a common longitude of
periastron (Ji \etal 2003, and references therein).

The four- or five-body system is integrated for 100 Myr or until the system
becomes unstable through a collision or an ejection.  We employ the hybrid
integrator in Mercury (Chambers 1999), which uses a second-order mixed variable
symplectic algorithm when objects are separated by more than 3 Hill radii, and
a Bulirsch-Stoer method for closer encounters.  The timestep in each system
was chosen in order to sample the
smallest orbit 20 times each period.  Our integrations typically conserved
energy to one part in 10$^5$.  Each simulation took zero to ten days to run on
a desktop PC, depending on the system and the outcome of the simulation (some
systems resulted in ejections or collisions within a few wall clock minutes).

\section{Results}

We present the results for three systems which were shown in Paper 1 to contain
zones which are stable for massless test particles for at least 5-10 Myr.  In
addition, we examine the system HD74156 which did not contain such a stable
zone.  In Table 2, we present the initial conditions for the simulations of
each system, including the parameter space sampled and the number of
Saturn-mass planet experiments.  Table 3 summarizes our results. 

\subsection{HD37124}

This system has an interesting resonant structure.  The ratio
of the periods of the two known giant planets is 12.7, making it by far the
most compact of our candidate systems.  Three mean-motion resonances with the
inner planet lie near the sampled region of parameter space -- the 2:1 (0.86
AU), 3:1 (1.12 AU), and 5:2 (0.995 AU).  The 5:2 resonance bisects the sampled
region and has important consequences for the survival rate of test planets,
as shown below. 

\begin{figure*}
\centerline{
\psfig{file=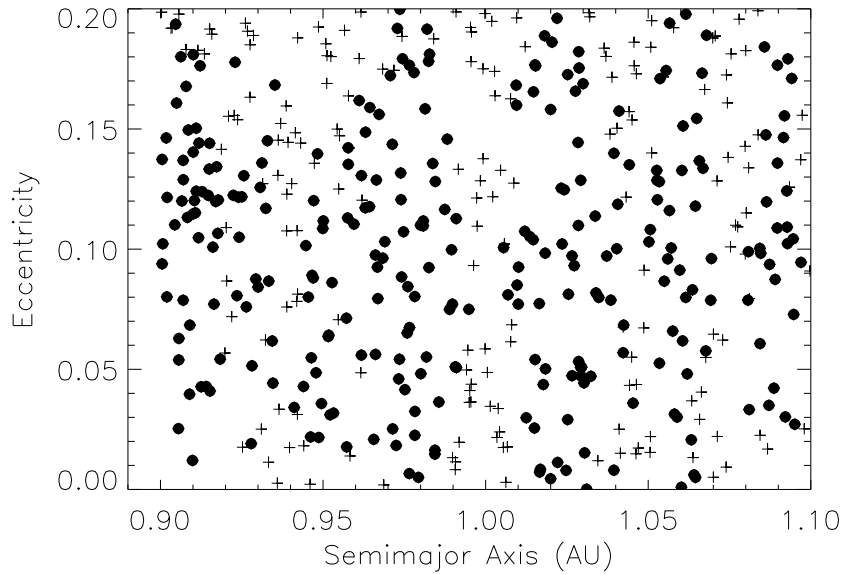,width=12cm}}
\caption{The distribution in ($a$, $e$) space of 472 Saturn-mass planets
in HD37124.  Solid dots represent systems which were stable for 100 Myr, and
crosses represent unstable configurations. }
\label{fig:37a}
\end{figure*}

Paper 1 showed that test particles are stable in HD37124
for semimajor axes between 0.9 and 1.1 AU, with eccentricities between 0 and
0.25.  We integrated the orbits of 472 Saturn-mass planets in this system, 290
(61\%) of which survived for 100 Myr.  Figure~\ref{fig:37a} shows the distribution in
($a$, $e$) space of the Saturns, in which solid dots represent planets which
survived and crosses represent unstable configurations.  

\begin{figure*}
\centerline{
\psfig{file=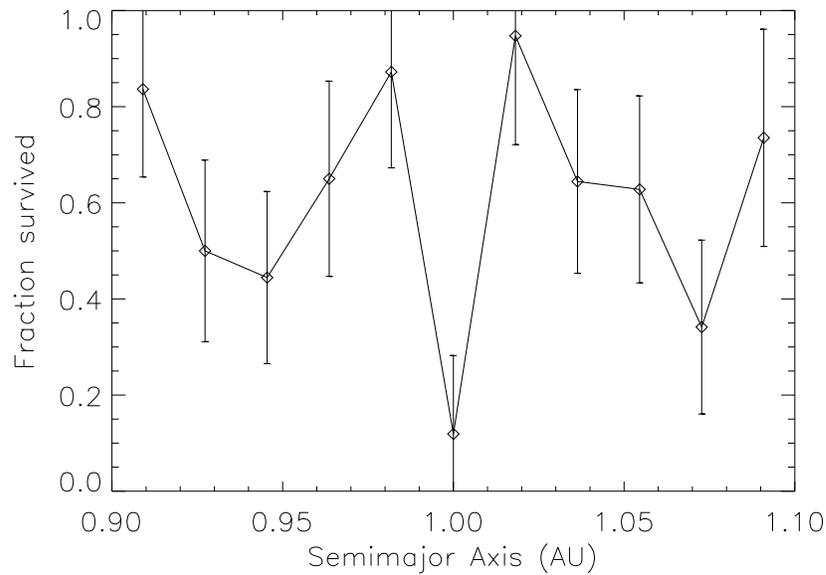,width=12cm}}
\caption{The survival rate of Saturn-mass planets in HD37124 as a function of
semimajor axis, with Poisson error bars.  Note the strong instability at the 2:5
mean motion resonance, and the stable regions immediately interior and exterior.}
\label{fig:37c}
\end{figure*}

Figure~\ref{fig:37c} shows the survival rate of Saturn-mass planets as a
function of semimajor axis including Poisson error bars.  Note the strong
decline in survival rate at the 2:5 mean motion resonance with the inner
planet (located at 0.995 AU), and the peaks in survival rate immediately
interior and exterior.  There is a smaller peak at $a \simeq$ 0.90 AU.  We saw
no strong dependence of the survival rate on mean anomaly in the resonance.
The peaks on either side of the resonance are reminiscent of Jupiter and
Saturn in our solar system, which lie slightly out of perfect 5:2 resonance,
and whose orbits are stable.  

\begin{figure*}
\centerline{
\psfig{file=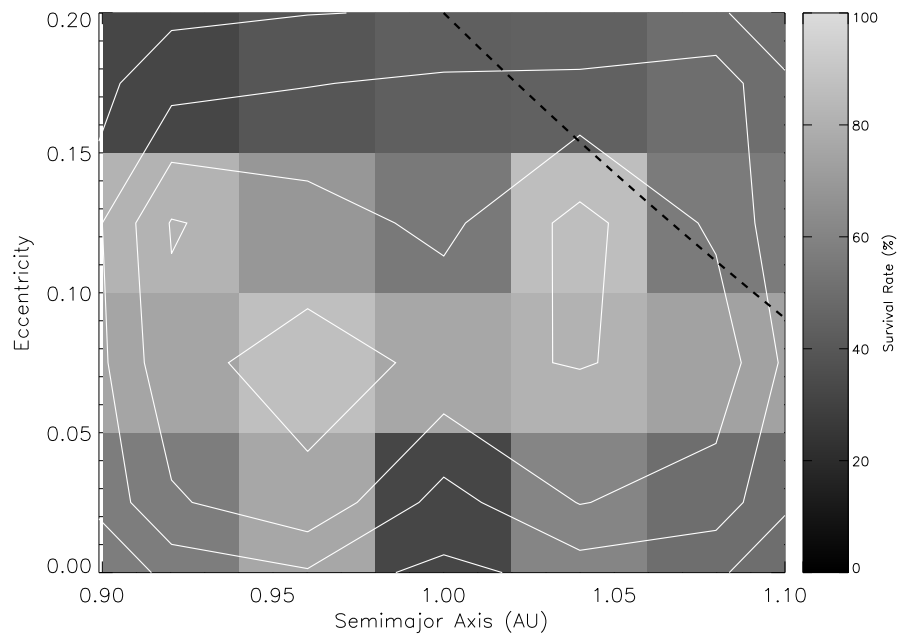,angle=90,width=12cm}}
\caption{The data for HD37124 from Fig.~\ref{fig:37a}, binned on the $a$
and $e$ axes.  The shade of
each bin represents the fraction of planets in that bin which survived for 100
Myr, with Poisson error of roughly 20\%.  Contours of constant survival
rate are over-plotted to bring out structure, spaced by 20\%.  The black
dashed line is the outer edge of the system's habitable zone.  Note the
three local maxima, including one on either side of the 2:5 resonance
at 0.995 AU.}
\label{fig:37b}
\end{figure*}

Figure~\ref{fig:37b} shows the data averaged into bins in both $a$ and $e$
such that each bin contains roughly 25 points.
The shade of each square represents the fraction of planets in that
bin which survived, and has a Poisson error of 20\%.
Over-plotted are contours of constant survival rate, also spaced by 20\%, to show
the underlying distribution and the location of maxima.  The outer edge of the
system's habitable zone is marked by the black dashed line.  Three local maxima
are evident in the Figure: 1) $a$ $\sim$ 0.92 AU, $e \sim$ 0.12, 2) $a
\sim$ 1.02 AU, $e$ $\sim$ 0.1, and 3) the absolute maximum at
$a$ $\sim$ 0.98 AU, $e$ $\sim$ 0.07.  Each of these maxima is located in the
habitable zone of the system, although maximum 2 is at the outer edge.

\begin{figure*}
\centerline{\psfig{file=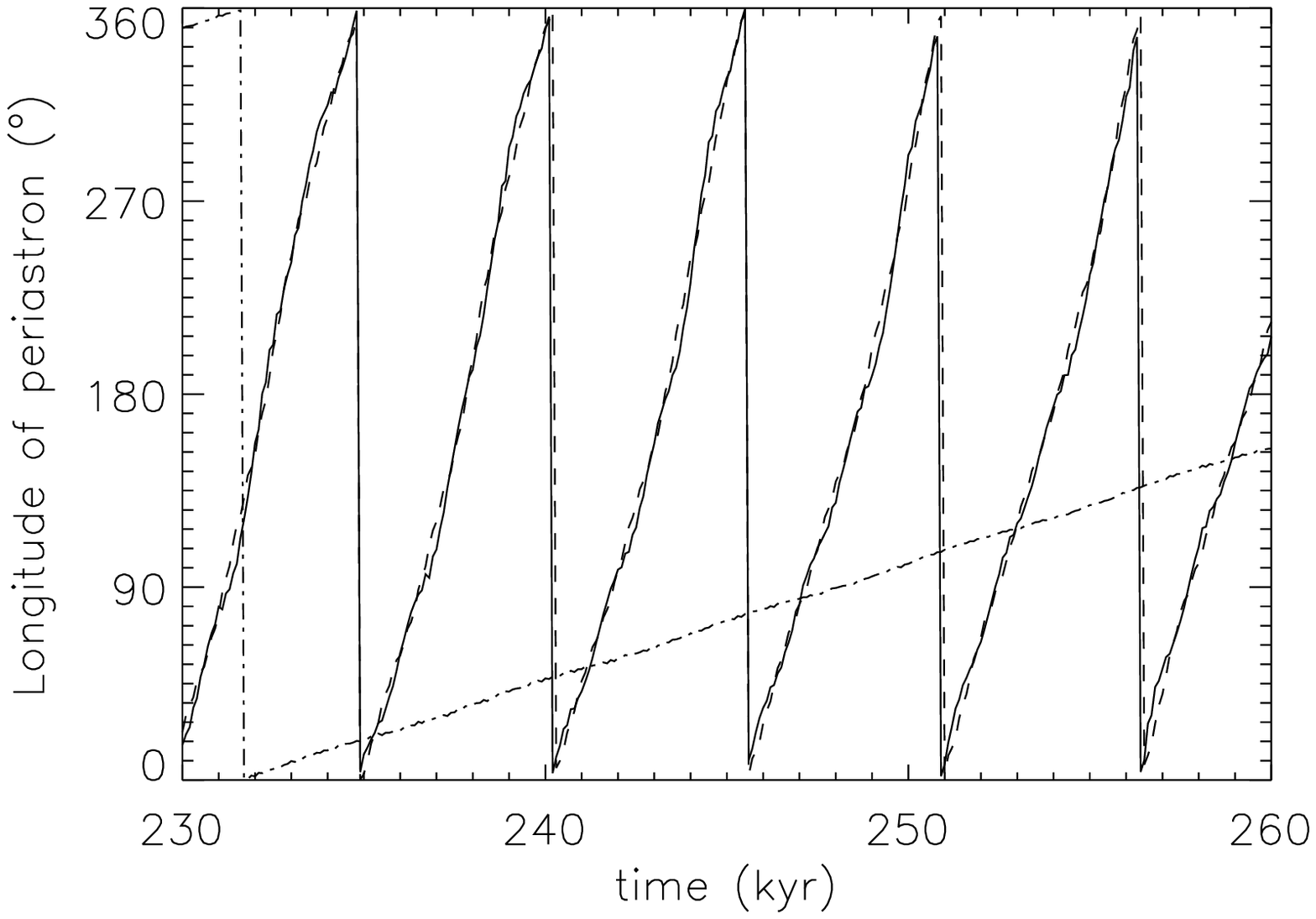,width=10cm}}
\centerline{\psfig{file=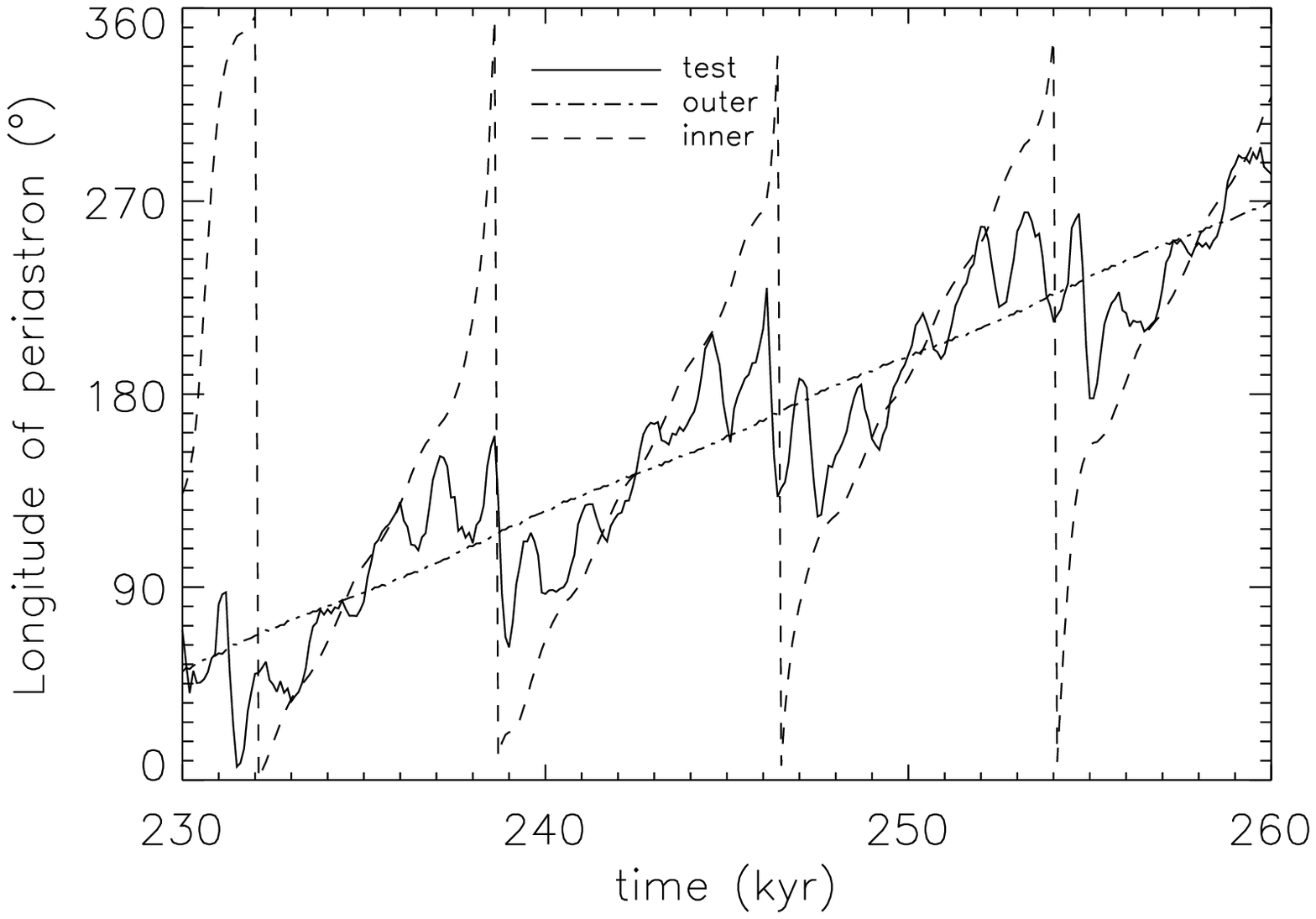,width=10cm}}
\caption{Evolution of the orientation of orbits (measured by the longitude of
periastron) for two test systems of HD37124.  The orbital elements of the
Saturn-mass test planets are ($a$, $e$) =  (0.90 AU, 0.11) (top)
and (1.01 AU, 0.08) (bottom).  Both systems were stable for 100 Myr.}
\label{fig:37lt}
\end{figure*}

In the absence of a test planet, the longitudes of periastron of the inner and
outer giant planet librate about each other with an amplitude of roughly
31$^{\circ}$ and a precession period of 171 kyr.  With the insertion of a
Saturn-mass test planet, we find evidence for secular resonances in various
configurations of test planets.  Fig.~\ref{fig:37lt} shows the time evolution
of the longitudes of periastron of each planet in two cases.  The initial orbital elements
for the test planet in these systems were ($a$, $e$) =  (0.90 AU, 0.11) (top
panel) and (1.01 AU, 0.08) (bottom panel).  The top panel shows a system in
which the Saturn-mass test planet is in a strong secular resonance with the
inner giant planet, as the orientation of the two planets' orbits are tracking
each other with time.  The bottom panel shows a case in which the test planet's
longitude of periastron is librating about that of the outer giant planet.  At
the same time, the test planet's orbit tracks that of the inner giant planet
for over half of its precession cycle of $\sim$ 7.5 kyr (e.g. 231 to 237 kyr).
An additional, 1.5 kyr oscillation is superimposed on the evolution of the
test planet.  The secular dynamics of the test planet in this case are
affected by both giant planets in a complex way, yet the system is
stable.  We expect simple systems in secular resonance to be stable because of
the avoidance of close approaches between planets.  The presence of the
Saturn-mass test planets makes an analytical treatment of the secular
resonance structure of the system beyond the scope of this
paper.  Note in Fig.~\ref{fig:37lt} that the precession rates of both the
inner and outer giant planets are different in the top and bottom panels, due
to the different locations of the test planet.

\subsection{HD38529}

The resonant structure of HD38529 is quite different from that
of HD37124, as the separation between the two known planets is much larger.
The stable region for test particles lies between 0.27 and 0.82 AU, with
eccentricities up to 0.3.  The inner edge is cut off by the 1:3 resonance with
the inner planet.  We see no evidence of secular resonances playing a
significant role in the dynamics.

\begin{figure*}
\centerline{
\psfig{file=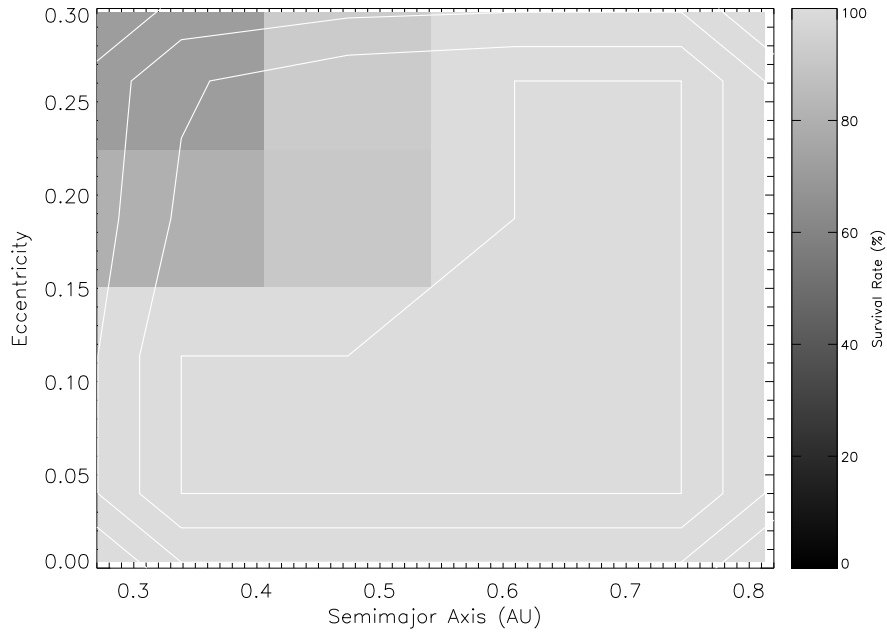,angle=90,width=12cm}}
\caption{Binned data from 200 simulations of Saturns in HD38529 with contours
of constant survival rate over-plotted, as in Fig.~\ref{fig:37b}.  Contours of
constant survival rate are spaced by 25\%.  The only
unstable systems lie at low $a$ and high $e$.}
\label{fig:38a}
\end{figure*}

We integrated the orbits of 200 Saturns in
this system, of which 191 (95.5\%) survived for 100 Myr.  Figure~\ref{fig:38a}
shows the data binned and over-plotted with contours as in Fig.~\ref{fig:37b}.
The only unstable regions in this system lie at small semimajor axes
and high eccentricities.  The vast majority of the zone which is stable for
massless test particles is also stable for Saturn-mass planets.  It is
therefore difficult to dynamically constrain the location of such a planet
beyond the results of Paper 1, although its orbit would likely be stable in
the given region. 

\subsection{55 Cancri} 

This system is interesting dynamically, as it is composed of
an interior pair of planets in 3:1 mean motion resonance with a distant,
separated companion.  Paper 1 showed that there exists a large region between the
inner pair and the outer planet which is stable for test particles, at 0.7 AU
$< a <$ 3.4 AU, with eccentricities up to 0.2.  This stable region is bounded at
its inner edge by the 1:5 resonance with the inner planet at 0.72 AU, and at
its outer edge by the 5:2 resonance with the outer planet at 3.2 AU.  Several
mean motion resonances with the outer planet are located in the stable region,
notably the 3:1 resonance at 2.84 AU, the 4:1 resonance at 2.34 AU, and the
5:1 resonance at 2.02 AU. 

\begin{figure*}
\centerline{
\psfig{file=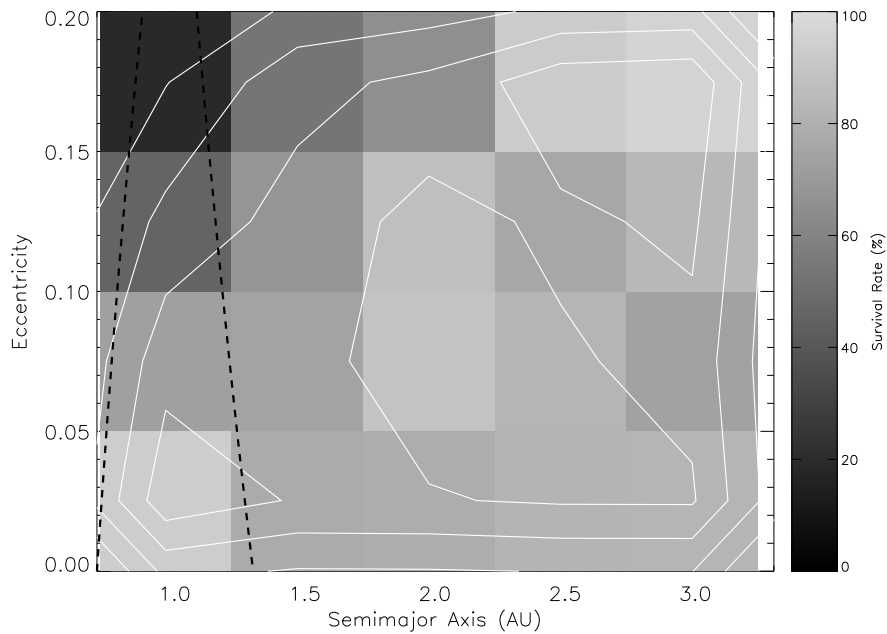,angle=90,width=12cm}}
\caption{Binned data from 512 simulations of Saturns in 55Cnc, with contours
of constant survival rate 
spaced by 20\%.  The black dashed lines indicate the boundaries of the
system's habitable zone.  Note the maxima at ($a$, $e$)
$\simeq$ (1.03 AU, 0.03), (2.0 AU, 0.08), and (3.0 AU, 0.17).}
\label{fig:55a}
\end{figure*}

We integrated the orbits of Saturns in 512
locations within this zone, and 384 (75\%) of these survived for 100 Myr.
Figure~\ref{fig:55a} shows the distribution in ($a$, $e$) space of our
experiments, in the same format as Figs.~\ref{fig:37b} and~\ref{fig:38a}, with
Poisson errors of $\sim$ 20\% per bin.
We see three local maxima: 1) a relatively narrow maximum at $a$ $\sim$ 1.0
AU, $e$ $\sim$ 0.03, 2) a broad maximum centered roughly at $a$ $\sim$ 2.0 AU,
$e$ $\sim$ 0.08 but which extends to higher values of $a$, and 3) $a$ $\sim$ 3
AU, $e$ $\sim$ 0.17.  Region 1 is of great astrobiological interest, as it
lies in the habitable zone of its parent star, which is bounded by the black
dashed lines. Region 3 is bordered
by the 3:1 (2.84 AU) and 5:2 (3.2 AU) mean motion resonances with the outer
planet.  We see no clear trend of survival rate with mean anomaly near these
resonances. 

\subsection{HD74156}

Paper 1 found that no test particles survived in this system for
longer than 1 Myr.  The region in which they survived the longest was for
$a$ between 0.5 and 1.5 AU at relatively low eccentricities.  The 1:5 mean
motion resonance (with the inner planet) is at 0.82 AU and the 5:1 resonance
(with the outer planet) at 1.3 AU are located at the outskirts of the region
we investigate.  Therefore, only very high order mean motion resonances are
found in the center.  We find no evidence of secular resonances in the region.  

\begin{figure*}
\centerline{
\psfig{file=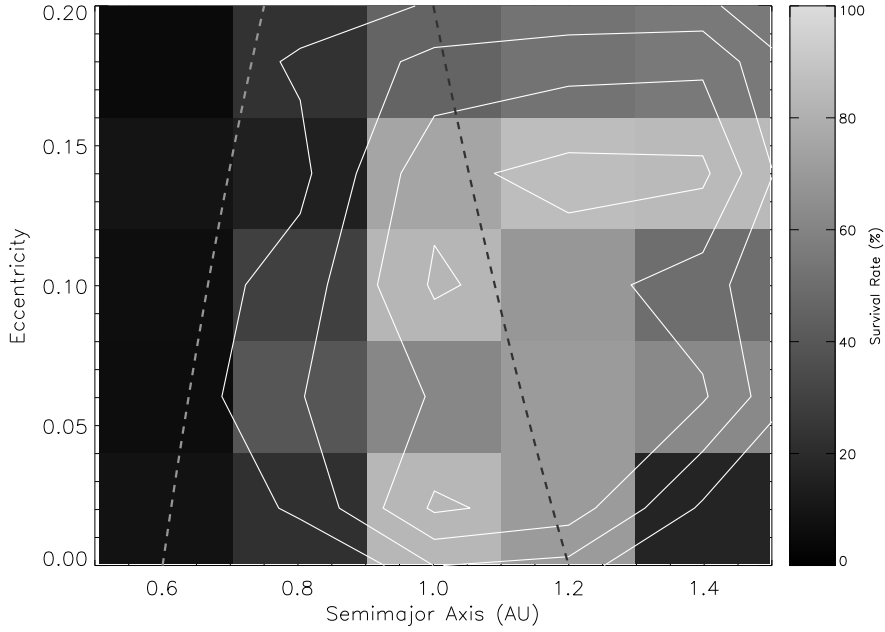,angle=90,width=12cm}}
\caption{Binned data from 600 simulations of Saturns in HD74156, formatted as
in Fig.~\ref{fig:37b}, with contours of constant survival rate
spaced by 20\%.   The dashed lines indicate the boundaries of the
system's habitable zone.  The absolute maximum is located at ($a,e$) $\simeq$ (1.0 AU,
0.02) and two local maximum are at (1.0 AU, 0.10) and (1.2 AU, 0.13).}

\label{fig:74a}
\end{figure*}

We performed
600 integrations of Saturn-mass planets in this system with $a$ in the
above mentioned region, and $e$ between 0 and 0.2.  Of these 600 Saturns, 296
(49\%) survived for 100 Myr.  Figure~\ref{fig:74a} shows the distribution of
the surviving planets in these simulations.  We see three small islands of
stability at ($a,e$) $\simeq$: 1) (1.0 AU, 0.02), 2) (1.0 AU, 0.1), and 3)
(1.2 AU, 0.13).  These three islands lie at a slightly higher survival rate
than the surrounding, larger region of stability between 0.9 - 1.2 AU with $e
\leq$ 0.15, in which the survival rate is 75\%.

\begin{figure*}
\centerline{
\psfig{file=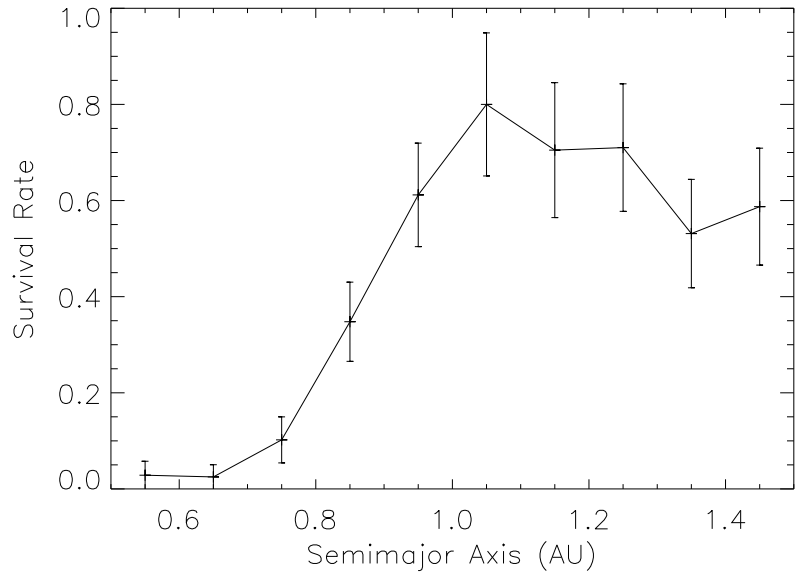,width=12cm}}
\caption{Survival rate of Saturns in HD74156 as a function of semimajor axis,
with statistical error bars.  Note the strong increase toward 1 AU and the
plateau between 1.0 and 1,3 AU.}
\label{fig:74b}
\end{figure*}

We see a strong trend in the survival rate of planets as a function of
semimajor axis, as shown in Fig.~\ref{fig:74b}.  The fraction of systems which
are stable for 100 Myr increases sharply between 0.8 and 1.0 AU, then flattens
off and decreases slightly past 1.2 AU.  The stable zones found in
Fig.~\ref{fig:74a} lie at the peak of the curve.

\section{Discussion}

Menou \& Tabachnik (2003; hereafter MT) investigated the possibility of
Earth-sized planets
residing in the habitable zones (HZs) of known extrasolar planetary systems.  The
location of the HZ is a function of the luminosity (and
therefore mass) of the host star, as well as the atmospheric composition of
the planet (Kasting \etal 1993).  For each system MT integrated the orbits
of 100 massless test particles in the HZ for 10$^6$ years.  They
considered all four of our systems.  The HZs for each system are
as follows -- HD37124: 0.6-1.2 AU, HD38529: 1.4-3 AU, HD74156: 0.6-1.2 AU, and
55Cnc: 0.7-1.3 AU.  MT found no surviving planets in the HZ of
HD37124.  Their stability criterion requires a particle to remain in the HZ
at all times, limiting its eccentricity such that the particle's aphelion and
perihelion remain in the HZ.  Paper 1 used over 500 test particles to
systematically map out the region in HD37124 which is stable for test
particles, finding it to be centered at 1 AU.  The eccentricities in this
stable region are small enough to keep test particles in the HZ of the
system throughout their orbits.
In addition, we find three local maxima of the survival rate Saturn-mass planets
in this system, all of whose orbits remain in the HZ.  

For HD38529 our
results are consistent with MT, as the stable region from Paper 1 lies well outside
the HZ, and the region we investigated with Saturns does not
overlap with the HZ.  In the case of 55Cnc our results are again
consistent with MT, who find that a significant fraction of low-inclination
test particles survive at 1.0 AU, with eccentricities centered on 0.09.  The
stable region for 55Cnc from Paper 1 encompasses the HZ entirely
for eccentricities below 0.25.  In addition, Table 3 shows a maximum in the
survival rate of Saturns at ($a$, $e$) = (1.0 AU, 0.03), very close
to the value from MT.
MT's results for HD74156 are consistent with Paper 1, but we have found two regions in
the HZ which are stable for Saturn-mass planets in 83\% of cases.
However, this may be due to the fact that the orbital elements used by MT are
different than those we have used here.  In particular, the semimajor
axis of the outer planet used here is 0.35 AU larger (3.82 AU vs 3.47 AU),
increasing the separation of the two giant planets and therefore possibly
causing the region in between to become more stable for an additional
companion.  Note that the current value for HD74156c is 
3.40 AU (Naef \etal 2004).  

Dvorak \etal (2003) investigated the possibility of an unseen planet in
HD74156, using both test particles and massive ones.  They find a broad,
relatively stable region for test particles between 0.9 and 1.4 AU, with the
most stable location being at $a$ = 1.25 AU and $e$ $<$ 0.2.  This is a region
in which Paper 1 found no stable test particle orbits.  Fig.~\ref{fig:74b} shows
a plateau in survivability between 1.0 and 1.25 AU.  Dvorak
\etal (2003) found no trend in the results of their simulations
of massive planets, and concluded that the presence of an unseen companion in
the system was unlikely.  Further observations will shed light on this issue,
although the 75\% survival rate of Saturns for the entire region with 0.9 AU
$< a <$ 1.2 AU, $e \leq$ 0.15 suggests that this is a real possibility.  Note again
that the best-fit orbit of the outer planet in this system has
recently been revised to $a$ = 3.40 AU, $e$ = 0.58 (Naef \etal 2004).  The
closer proximity and higher eccentricity of this planet strongly affects the
dynamics between the two known plants.  Both Dvorak \etal (2003) and
Paper I assume the orbital elements from Table 1 in their calculations.

\section{Conclusions}

We have found specific locations in four known extrasolar planetary systems in
which Saturn-mass planets could exist on stable orbits.  Such a planet would
lie just below the detection threshold of current radial velocity surveys, and
may be detected in the near future.  Table 3 summarizes our results, detailing
the location in ($a,e$) space of each maximum in the survival rate for each of
our four candidate systems.  If an additional planet is discovered in the
stable region of one of these systems, it would mark the first successful
prediction of a planet since John Couch Adams predicted the existence of
Neptune in 1845 based on perturbations to Uranus' orbit. 

Does the presence of a stable region imply the presence of a planet?  Must
all systems contain as many planets as they can?  Laskar (1996) speculated
that ``a planetary system will always be in this state of marginal stability,
as a result of its gravitational interactions.''  The ``packed planetary
systems'' (PPS) hypothesis, presented in Paper 1 (see also Barnes \& Quinn,
2004), extends this idea by
suggesting that all systems contain as many planets as they can dynamically
support without self-disrupting.  
All systems may be on the edge of stability, but observational constraints
prevent the detection of smaller or more distant bodies which push apparently
stable systems to this edge. 

The formation scenario of a planet of any size in between two gas giant
planets is of great interest.  In the Solar System no stable regions exist
between the orbits of the gas giants.  The detailed formation scenario of a smaller
giant planet between two others is unclear, be it through gravitational
instability (e.g. Mayer \etal 2002) or core-accretion (Pollack \etal 1996).  Gas
giant planets at small orbital radii may have formed farther out in the
protoplanetary disk and migrated inward, which further complicates this
formation scenario.  

Certain stable regions in HD37124, 55Cnc and HD74156 are located in the
habitable zones of their parent stars (see Table 3).  Clearly, the discovery
of a planet of any size in these regions is of great astrobiological importance,
as any giant planet would likely have one or more large moons.  Understanding
the formation of terrestrial planets in these systems is vital. 
In the upcoming third paper of the ``predicting planets'' series (Raymond \&
Barnes 2004) we present results of simulations of terrestrial planet formation
in between the known giant planets in the same four systems examined here.  

\section{Acknowledgments}
We thank Tom Quinn and Andrew West for many helpful discussions, and Chance
Reschke for his assistance in the completion of
the simulations presented in this paper. This work was funded
by grants from the NASA Astrobiology Institute, the NSF, and a NASA
GSRP. These simulations were performed on computers donated by the University
of Washington Student Technology Fund. These simulations were performed under
CONDOR.\footnote{CONDOR is publicly available at http://www.cs.wisc.edu/condor}

\scriptsize
\begin{deluxetable}{c|c|c|c|c|c|c}
\tablewidth{0pt}
\tablecaption{Orbital Parameters of Selected Planetary Systems}
\renewcommand{\arraystretch}{.6}
\tablehead{
\\
\colhead{System} &  
\colhead{Planet} & 
\colhead{M ($M_J$)} &
\colhead{$a$ (AU)} &  
\colhead{$e$}&
\colhead{$\varpi$}&
\colhead{T (JD)}}
\startdata

\\
HD37124 & b & 0.86 & 0.54 & 0.1 & 97.0 & 2451227\\
 & c & 1.01 & 2.95 & 0.4 & 265.0 & 2451828\\
\hline
\\
HD38529 & b & 0.78 & 0.129 & 0.29 & 87.7 & 2450005.8\\
 & c & 12.8 & 3.68 & 0.36 & 14.7 & 2450073.8\\
\hline
\\
55Cnc & b & 0.84 & 0.115 & 0.02 & 99.0 & 2450001.479\\
 & c & 0.21 & 0.241 & 0.339 & 61.0 & 2450031.4\\
 & d & 4.05 & 5.9 & 0.16 & 201.0 & 2452785\\
\hline
\\
HD74156\tablenotemark{1} & b & 1.61 & 0.28 & 0.647 & 185.0 & 2451981.38\\
 & c & 8.21 & 3.82 & 0.354 & 272.0 & 2451012.0\\
\enddata
\tablenotetext{1}{Best fit values as of August 22, 2002.  The current best fit
 for planet c is $a$ = 3.40 AU, $e$ = 0.58 (Naef \etal 2003).}
\end{deluxetable}

\scriptsize
\begin{deluxetable}{c|c|c|c}
\tablewidth{0pt}
\tablecaption{Initial Conditions for Simulations}
\renewcommand{\arraystretch}{.6}
\tablehead{
\\
\colhead{System} &  
\colhead{$\Delta a$ (AU)} &
\colhead{$\Delta e$}&
\colhead{N (Saturns)}}
\startdata
\\
HD37124 & 0.9 -- 1.1 & 0.0 -- 0.2 & 472\\
\hline
\\
HD38529 & 0.27 -- 0.82 & 0.0 -- 0.3 & 200\\
\hline
\\
55Cnc & 0.7 -- 3.2 & 0.0 -- 0.2 & 512\\
\hline
\\
HD74156\tablenotemark{1} & 0.5 -- 1.5 & 0.0 -- 0.2 & 600\\
\enddata
\tablenotetext{1}{Paper 1 found that no test particles in HD74156 survived for
longer than 1 Myr.  In our simulations, however, we sample the given region of
parameter space.}
\end{deluxetable}

\scriptsize
\begin{deluxetable}{c|c|c}
\tablewidth{0pt}
\tablecaption{Simulation Results}
\renewcommand{\arraystretch}{.6}
\tablehead{
\\
\colhead{System} &  
\colhead{Stable Region ($a$,$e$)\tablenotemark{1}} &
\colhead{Survival Rate\tablenotemark{2}}}
\startdata
\\
HD37124  & (0.92 AU, 0.12)\tablenotemark{*} & 81\%\\
         & (0.98 AU, 0.07)\tablenotemark{*} & 87\%\\
	 & (1.02 AU, 0.1)\tablenotemark{*} & 84\%\\
	
\hline
\\
HD38529 & (0.3-0.8 AU, 0.0-0.15) & 100\%\\
\hline
\\
55Cnc & (1.0 AU, 0.03)\tablenotemark{*} & 93\%\\
      & (2.0 AU, 0.08) & 89\%\\
      & (3.0 AU, 0.17) & 96\%\\
\hline
\\
HD74156 & (1.0 AU, 0.02)\tablenotemark{*} & 83\%\\
        & (1.0 AU, 0.10)\tablenotemark{*} & 83\%\\
        & (1.2 AU, 0.13) & 86\%\\
\enddata
\tablenotetext{1}{Local maxima of the survival rate, i.e. the center of each
        bin from Figs~\ref{fig:37b},~\ref{fig:38a},~\ref{fig:55a}, and~\ref{fig:74a} in
        which the survival rate is a maximum.  The exact location of the
        stable region is uncertain on the order of the bin size.} 
\tablenotetext{2}{Survival Rate for all simulations in the binned region in
which the stable region is located.  See
Figs~\ref{fig:37b},~\ref{fig:38a},~\ref{fig:55a}, and~\ref{fig:74a}. }
\tablenotetext{*}{Stable regions which lie in the habitable zone of their
parent stars, as defined by Kasting \etal (1993).}
\end{deluxetable}

\end{document}